\documentclass[twocolumn,showpacs,preprintnumbers,amsmath,amssymb]{revtex4}
\usepackage{graphicx}% Include figure files
\usepackage{dcolumn}% Align table columns on decimal point
\usepackage{bm}% bold math

\newcommand{\bi}{\bigskip}
\newcommand{\no}{\noindent}
\newcommand{\be}{\begin{eqnarray}}
\newcommand{\ee}{\end{eqnarray}}
\newcommand{\hk}{\hspace{0.1cm}}

\newcommand{\rk}{\right)}
\newcommand{\lk}{\left(}

\begin{document}

\preprint{}

\title{Quark and gluon confinement in Coulomb gauge}% Force line breaks with \\

\author{C. Feuchter and H. Reinhardt}
 \affiliation{Institut f\"ur Theoretische Physik\\
Auf der Morgenstelle 14\\
D-72076 T\"ubingen\\
Germany}%Lines break automatically or can be forced with \\
%\affiliation{%

%\affiliation{

\date{\today}% It is always \today, today,

\begin{abstract}
The Yang-Mills Schr\"odinger equation is variationally solved in Coulomb gauge 
for the vacuum sector using a trial wave functional, which is
strongly peaked at the Gribov horizon. We find the absence of gluons in the
infrared and also a confining quark potential.
\end{abstract}

\maketitle

%\section{\label{sec:level1}Introduction}
\no
{\it 1. Introduction.}
One of the most
challenging problem in particle physics is to explain the confinement of quarks
and gluons in QCD. 
Several confinement mechanisms have been proposed in the past, the
most prominent of which are perhaps the condensation of magnetic monopoles 
(dual Meissner effect) and center vortex
condensation (for a recent review see ref.\cite{R1}). Evidence for
both mechanisms has been found in lattice calculations.  
However, the center vortex picture is perhaps more
advantageous in the sense, that these vortices can, in principle, be defined in a gauge
invariant way \cite{R2}, while magnetic monopoles
arise as gauge artifacts after Abelian projection and are merely manisfestations
of topological defects of the underlying gauge fields \cite{R3}. 
Yet another confinement mechanism was proposed by 
Gribov \cite{R5} and further elaborated in ref.\cite{R6}. This mechanism,
which is
based on the infrared dominance of the field configurations near the Gribov
horizon in Coulomb gauge, is compatible with the magnetic monopole and 
center vortex
pictures of confinement, 
given the fact, that these field configurations lie on the
Gribov horizon \cite{R7}. 
%\bigskip

\no
In this letter we will explore Gribov's confinement
 mechanism by
studying the vacuum sector of $SU(2)$ Yang-Mills theory in Coulomb gauge in the Hamilton
approach using the variational principle. 
With an appropriate, physically motivated
ansatz for the Yang-Mills wave functional, which accounts for the dominance of
the field configurations on the Gribov horizon, we will be able to describe
simultanously quark and gluon confinement. \newline
In previous studies of the Yang-Mills Schr\"odinger equation in Coulomb gauge 
\cite{R8,8A}, a different trial wave
function was used and also the non-trivial metric of the orbit space, induced by
the Faddeev-Popov determinant was not fully included. In ref.\cite{R8} the
curvature of orbit space was completely neglected, while in ref.\cite{8A} its
contribution to the gluon self-energy was partially omitted in the gap equation. We will
find, however, that the proper inclusion of the Faddeev-Popov determinant is
absolutely necessary to produce simultaneously quark and gluon confinement.
%\vspace{0.1cm}

\no
{\it 2. Yang-Mills theory in Coulomb gauge.}
The Hamilton approach to gauge theory is based on the Weyl gauge $A_0 = 0$, in
which the dynamical degrees of freedom are the spatial components of the gauge
field, $\vec{A} (x)$. This gauge leaves still invariance under spatial gauge
transformations. The latter can be fixed by implementing the Coulomb gauge
$\vec{\partial} \vec{A} = 0$. In the Coulomb gauge the unphysical longitudinal
degrees of freedom of $\vec{A} (x)$ can be completely eliminated by explicitly 
resolving
Gau\ss' law $\hat{D} \Pi | \Psi \rangle = \rho_{ext}| \Psi \rangle $, 
resulting in the following Hamiltonian of the physical
transversal degrees of freedom $A^\perp_i$ $(\partial_i A^\perp_i = 0)$
\begin{widetext}
\be
\label{1}
H & = & \frac{1}{2} \int d^3 x \left[ g^2 {\cal{J}}^{- 1} [A^\perp] \Pi^a_i (x) 
{\cal{J}} [A^\perp]
\Pi^a_i (x) + \frac{1}{g^2} (B^a_i (x))^2 \right] \nonumber\\
& & + \frac{g^2}{2} \int d^3 x \int d^3 x' {\cal{J}}^{- 1} [A^\perp] \rho^a (x) \langle
a, x | (- \vec{\partial} \vec{\hat{D}})^{- 1} (- \partial^2) (- \vec{\partial}
\vec{\hat{D}})^{- 1}  | b, x' \rangle {\cal{J}} [A^\perp] \rho^b (x')
\hk .
\ee
\end{widetext}
Here $g$ is the Yang-Mills coupling constant,
$\Pi^a_i (x) = \frac{1}{i} \delta / \delta A^\perp_i (x)$ is the
canonical momentum operator conjugate to the transversal gauge field
$\vec{A}^\perp
(x)$, $\hat{D} = \partial + \hat{A}^\perp$ is the covariant derivative in the
adjoint representation ($\hat{A}^{a b} = f^{a c b} A^c , f^{a b c}$ being the
structure constant) and ${\cal{J}} = Det ( - \vec{\partial} \vec{\hat{D}})$ is the Faddeev-Popov
determinant. Given the fact, that the Faddeev-Popov kernel $(- \vec{\partial}
\vec{\hat{D}})$ represents the metric tensor in the color space of transversal
gauge connections $A^\perp_i$ the first term in the Hamiltonian is the
corresponding Laplace-Beltrami operator and gives the electric part of the
Hamiltonian. The second term gives the magnetic energy with $B_i [A^\perp] = \frac{1}{2}
\epsilon_{i j k} [D_j, D_k]$ being the color magnetic field. This term
represents a potential for the transversal gauge field. Finally the last term is
the so-called Coulomb term, where $\rho^a = \hat{A}_i^{\perp^{a b}}
\Pi^b_i + \rho^a_{ext}$ is the non-Abelian color charge density, which
contains besides the gluonic part also a contribution from external quarks
$\rho_{ext}$. \newline
%\bi
%\no
In this letter we solve the Yang-Mills Schr\"odinger equation $H \Psi = E \Psi$
for the vacuum sector by the variational principle using the following ansatz
for the Yang-Mills vacuum wave functional
%\begin{widetext}
\be
\label{2}
 \Psi [A^\perp] = \frac{{\cal N}}{\sqrt{{\cal{J}}[A^\perp]}} \hk e^{ 
{- \frac{1}{2g^2} \int d^3 x  d^3 x'
{A^\perp}^a_i (x) \omega (x, x') {A^\perp}^a_i (x')} \hk} 
\ee
%\end{widetext}
where ${\cal N}$ is a normalization constant and 
the kernel $\omega (x, x')$ is determined by minimizing the energy. %\newline
%\bi
%\no
The wave functional (\ref{2}) is strongly peaked at the Gribov horizon,
where the Faddeev-Popov determinant vanishes and thus reflects the fact, that
the dominant infrared configurations, like center vortices or magnetic monopoles,
 lie on the Gribov horizon
\cite{R7}. Furthermore, the wave functional (\ref{2}), being divergent on the
Gribov horizon, identifies all configurations on the Gribov horizon, in
particular those which are gauge copies of the same orbit. This identification
is absolutely necessary to preserve gauge invariance. In addition it
topologically compactifies the (first) Gribov region. Thus the pre-exponential
factor ${\cal{J}}^{- \frac{1}{2}} [A]$ drastically changes the properties of the vacuum
state compared to those of a pure Gaussian, which was used in refs.\cite{R8},
\cite{8A}. In
the case of QED, where the Faddeev-Popov determinant becomes a constant 
eq.(\ref{2}) represents the exact vacuum wave functional with $\omega (k) =
\sqrt{\vec{k}^2}$ being the energy of a free photon with 3-momentum
$\vec{k}$.
%\vspace{-0.1cm}
\bi

\begin{figure}
\includegraphics [scale=0.5] {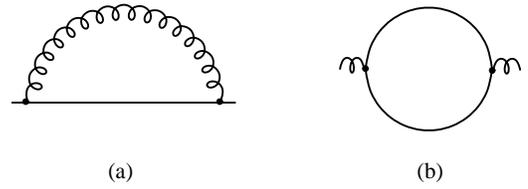}% Here is how to import EPS art
\caption{\label{diagram1u2} (a) The ghost self-energy in the rainbow ladder approximation. The full
(wavy) line represents the full ghost (gluon) propagator. (b) Ghost loop contribution to
the gluon self-energy.}
\end{figure}

\no
{\it 3. Schwinger-Dyson equations.}
In the evaluation of the vacuum energy $\langle \Psi | H | \Psi \rangle = \int D
A^\perp {\cal{J}} [A^\perp] \Psi^* [A^\perp] H \Psi [A^\perp]$ the
following ingredients are required:

\no
1. The ghost propagator in the vacuum, which is defined by
\be
\label{3}
G = \langle \Psi 
| ( - \vec{\partial} \vec{\hat{D}})^{- 1} | \Psi \rangle = (-
\vec{\partial}^2)^{- 1} \frac{d}{g} \hk ,
\ee
where $d$ denotes the ghost form factor, which measures the deviations of the
ghost from a free massless field.
Evaluating this expectation value in the so-called rainbow ladder approximation,
where the self-energy of the ghost is given by the diagram shown in 
fig.\ref{diagram1u2}a one obtains the following integral equation 
for the ghost form factor $d$ in momentum space
%\vspace{-0.2cm}

%\vspace{1.0cm}
%[scale=0.3,bb=15 554 470 758,clip=]
%\begin{figure}
%\centerline{
%\label{F1}
%\epsfysize=1.0cm
%\epsffile{F1.eps}
%}
%\caption{The ghost self-energy in the rainbow ladder approximation. The full
%(wavy) line represents the full ghost (gluon) propagator.}
%\end{figure}
\be
\label{5}
\frac{1}{d (k)} = \frac{1}{g} - I_d (k) \hk ,
\ee
where ($N_C$ number of colors)
\be
\label{6}
I_d (k) = \frac{N_C}{2} \int \frac{d^3 q}{(2 \pi)^3} \lk 1 - (\hat{k}
\hat{q})^2 \rk \frac{d (k - q)}{(k- q)^2 {\omega (q)}} \hk .
\ee
%with $\hat{k}=\frac{\vec{k}}{k}$.
%\vspace{0.1cm}
% -----------------------------------------------------------------------------------------------
% -----------------------------------------------------------------------------------------------
%
2. The Coulomb form factor $f$ defined by
\be
\label{7}
\langle \Psi | (- \vec{\partial} \vec{\hat{D}})^{- 1} (- \Delta) (-
\vec{\partial} \vec{\hat{D}})^{- 1} | \Psi \rangle := G (- \Delta) f G \hk .
\ee
The calculation of this form factor consistent with the evaluation of the ghost
form factor yields the following integral equation
\be
\label{8}
f(k) = 1 + I_f (k) \hk ,
\ee
%\begin{widetext}
\be
\label{8.1}
I_f (k) =  \frac{N_C}{2} \int d^3 q \lk 1 - (\hat{k} \hat{q})^2 \rk
\frac{d^2 (k - q) f (k - q)}{(k- q)^2 \omega (q)} \hk ,
\ee
%\end{widetext}
which is diagrammatically illustrated in fig.2. In this equation
we will replace the full ghost form factor by its bare value $d(k)=1$, 
in order to carry out the calculations consistently to 1-loop order.
% -----------------------------------------------------------------------------------------------
\begin{figure}
\includegraphics [scale=0.45] {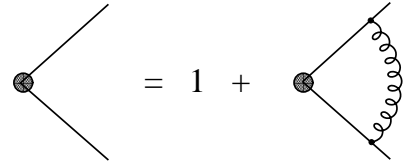}% Here is how to import EPS art
\caption{\label{diagram3} Diagrammatic illustration of the integral equation for the Coulomb form
factor $f$ represented by the fat dot with two entries for external legs.}
\end{figure}
% -----------------------------------------------------------------------------------------------
%\begin{figure}
%\centerline{
%\label{F2a}
%\epsfysize=2.0cm
%\epsffile{F2a.eps}
%}
%\caption{Diagrammatic illustration of the integral equation for the Coulomb form
%factor $f$ represented by the fat dot with two entries for external legs.}
%\end{figure}

3. The scalar ``curvature'' of color orbit space
%\begin{widetext}
\be
\label{9}
\chi (x, y) = -\frac{1}{4} \frac{1}{N^2_C - 1} \langle \Psi |
\frac{\delta^2 \ln {\cal{J}} [A^\perp]}{\delta {A^\perp}^{a}_i (x) \delta
{A^\perp}^{a}_i (y)} | \Psi \rangle \hk ,
\ee 
%\end{widetext}
which is the ghost loop part of the gluon self-energy and represents 
the ghost part of the color dielectric susceptibility of the
Yang-Mills vacuum.
To 1-loop order this curvature is given by the diagram shown in fig.1b
%\ref{F2}.
%\begin{figure}
%\includegraphics{F2}% Here is how to import EPS art
%\caption{\label{F2} Ghost loop contribution to the gluon self-energy}
%\end{figure}
%\begin{figure}
%\centerline{
%\label{F2}
%\epsfysize=2.0cm
%\epsffile{F2.eps}
%}
%\caption{Ghost loop contribution to the gluon self-energy}
%\end{figure}
and equals 
%\begin{widetext}
\be
\label{11}
\chi (k) & = & I_\chi (k) \hk , \nonumber \\
 I_\chi (k) 
& = & \frac{N_C}{4} \int \frac{d^3 q}{(2 \pi)^3} \lk 1 - (\hat{k}
\hat{q})^2 \rk \frac{d (k - q) d (q)}{(k- q)^2} \hk .\hk \hk \hk \hk
\ee
%\end{widetext}
%\end{enumerate}
%\bi

\no
Evaluating the expectation value of the Hamiltonian (\ref{1}) with the trial
wave functional (\ref{2}) to 1-loop order variation of the energy leads to
the following gap equation for the kernel $\omega (x, x')$ in the wave function
(\ref{2})
\be
\label{12}
\omega^2 (k) = k^2 + \chi^2 (k) + I^{(2)}_\omega (k) + 2 \chi (k) I^{(1)}_\omega (k) 
+ I^0_\omega \hk , \hk \hk
\ee
where $I^0_\omega$ is an irrelevant constant, which drops out after renormalization, and
%\be
%\label{13}
%I^0_\omega & = & \frac{N_C g^2}{4} \int \frac{d^3 q}{(2 \pi)^3} (3 - (\hat{k}
%\cdot \hat{q})^2) \frac{1}{\omega (q)} \hk , 
%\ee
%\begin{widetext}
\be
\label{14}
I^{(n)}_\omega (k) & = & \frac{N_C}{4} \int \frac{d^3 q}{(2 \pi)^3} \lk 1 + (\hat{k}
\hat{q})^2 \rk 
\cdot \frac{d (k - q)^2 f (k - q)}{(k - q)^2} \nonumber\\
& & \cdot \frac{(\omega (q) - \chi
(q) )^n - (\omega (k) - \chi(k) )^n }{\omega (q)} \hk .
\ee
%\end{widetext}
%\bi

\no
Equations (\ref{5}), (\ref{8}), (\ref{11}) and (\ref{12}) represent four coupled
Schwinger-Dyson type of equations for the ghost form factor $d (k)$, the Coulomb
form factor $f (k)$, the curvature $\chi (k)$ and the gluon energy $\omega (k)$.
These equations contain divergent integrals and require thus regularization and
renormalization. Fortunately, the asymptotic infrared and ultraviolet behaviour
of the solutions does not depend on the details of the renormalization procedure
used.
\bi

\no
{\it 4. Asymptotic behaviour.}
One can solve the coupled Schwinger-Dyson equations analytically in the
ultraviolet $k \to \infty$ in the so-called angular approximation \cite{RX3}. 
 One finds then the following
asymptotic ultraviolet behaviour
\be
\label{15}
\omega (k) \to \sqrt{\vec{k}^2} & , & \frac{\chi (k)}{\omega (k)} \to
\frac{1}{\sqrt{\ln k / \mu}} \\
\label{16}
d (k) \to \frac{1}{\sqrt{\ln k / \mu}} & , & f (k) \to
\frac{1}{\sqrt{\ln k / \mu}} \hk ,
\ee
where $\mu$ is an arbitrary parameter
of dimension mass. The first equation means, that the gluons behave asymptotically like free
particles, while the second one implies that the space of gauge connections
becomes asymptotically flat. The ghost and Coulomb form factors, $d (k)$ and
 $f (k)$,
deviate asymptotically from that of a free massless field by the anomalous
dimension factor $1 /\sqrt{\ln k /  \mu}$. \newline
%\bi
%\no
In the infrared one can rigorously show, that $\chi (k \to 0) = \omega (k \to
0)$ \cite{R10}. In addition, adopting the angular approximation \cite{RX3} 
one finds the asymptotic solution for $k \to 0$
\be
\label{16a}
\omega (k) = \chi (k) \sim \frac{1}{k} \hk , \hk d (k) \sim \frac{1}{k} \hk ,
\hk f (k \to 0) = 1 \hk .
\ee
The first relation implies, that in the infrared the gluon energy diverges and
equals its self-energy part generated by the ghost loop, i.e. its free part
$\sqrt{\vec{k}^2}$ has dropped out. The infrared diverging gluon energy is a
manifestation of gluon confinement. It implies an infrared vanishing gluon
propagator, which violates positivity, and accordingly the gluons do not occur
as asymptotic particle states in $S$-matrix \cite{RX4} ,\footnote{An infrared
vanishing gluon propagator and an infrared singular ghost propagator are also
found in the covariant Schwinger-Dyson approach in Landau gauge \cite{XY}.}. 
Furthermore, the infrared diverging ghost
form factor and the infrared finite Coulomb form factor $f (k \to 0) = 1$ is
precisely the infrared behaviour needed to produce a linearly rising confining
potential.
\bi

\no
{\it 5. Renormalization.}
To regularize and renormalize the divergent 
Schwinger-Dyson equations we use a 3-momentum cut-off and a
momentum substraction scheme similar to the one used in refs.\cite{R8,
8A}. However,
in the present case new features arises in the renormalization of the 
gap equation (\ref{12}) due to
the full inclusion of the curvature (\ref{9}). The details will be given
elsewhere \cite{R10}.  \newline
%\bi
%\no
After renormalization one obtains the following set of Schwinger-Dyson equations
\be
\label{17}
\frac{1}{d (k)} & = & \frac{1}{d (\mu)} - \Delta I_d (k) \\
\label{18}
\chi (k) & = & \chi (\mu) + \Delta I_\chi (k) \\
\label{XX}
f (k) & = & f (\mu) + \Delta I_f (k)
\ee
%\begin{widetext}
\be
\label{20}
\omega^2 (k) & = & k^2 - \mu^2 + (\Delta I_\chi (k))^2  
+ \xi \Delta I_\chi (k) + \Delta I^{(2)}_\omega (k) \nonumber\\
& &  + 2 [ \chi (\mu) + \Delta I_\chi (k) ] \Delta
I^{(1)}_\omega (k) + \omega ^2 (\mu)  ,
\ee
%\end{widetext}
where $\mu$ is the renormalization scale 
 and we have introduced the
abbreviations 
\be
\label{21}
\Delta I_d (k) = \lim\limits_{\Lambda \to \infty} \lk I_d (k, \Lambda) - I_d
(\mu, \Lambda) \rk \hk \hk \mbox{etc} \hk .
\ee
Furthermore $\omega (\mu), d (\mu), \chi (\mu), f (\mu)$ and $\xi = 2 [ \chi (\mu) + I^{(1)}_\omega (\mu)]$ are
renormalization constants, which were determined as follows: $\omega (\mu)$ is
used to fix the energy scale and drops out from the Schwinger-Dyson equations by
rewriting the latter in terms of dimensionless quantities. $d (\mu)$ enters only
the Schwinger-Dyson equation (\ref{17}) 
for the ghost form factor, which does not contain
the remaining renormalization constants. The ultraviolet behaviour of $d
(k)$ found above in eq.(\ref{16}) is independent of $d (\mu)$ but the infrared
behaviour of $d (k)$ 
depends crucially on $d (\mu)$. As long as $d (\mu)$ is smaller than
some critical value $d_{cr}$, $d (k)$ approaches a (finite) constant for $k \to
0$. At the critical value $d (\mu) = d_{cr}$ the ghost form factor $d (k)$
diverges for $k \to 0$ and above the critical value $d (\mu) > d_{cr}$ no
solution for $d (k)$ exists. This critical value is the
only value, which produces the infrared diverging ghost form factor found above
analytically (\ref{16a}). Furthermore in $D = 3$ (which will be considered elsewhere) a
self-consistent solution to the coupled Schwinger-Dyson equations exists only
for  this critical value. Therefore we choose $d (\mu) = d_{cr}$. Fortunately, the
self-consistent solution is quite insensitive to the remaining renormalization
constants $\chi (\mu)$ and $\xi$, which we have chosen for the definiteness
as $\chi (\mu) = 0, f (\mu) = 1$ and $\xi = 0$.
\bi

\no
{\it 6. Numerical Results.}
The renormalized coupled Schwinger-Dyson equations 
eq.(\ref{17}),(\ref{18}),(\ref{XX}) and (\ref{20}) are solved by iteration 
(without resorting to the angular approximation)
and the results are shown in
figs.3 and 4.
\begin{figure}
\includegraphics[scale=0.4,bb=42 34 600 406,clip=]{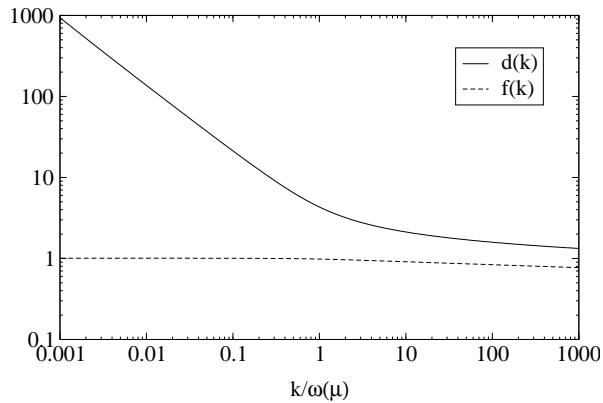}
\caption{\label{figure5F4} The ghost form factor $d (k)$ (full line) and the Coulomb form factor
$f (k)$ (dashed line).}
\end{figure}
%\begin{figure}
%\centerline{
%\label{5F4}
%\epsfysize=7.0cm
%\epsffile{figure5F4.eps}
%}
%\caption{The ghost form factor $d (k)$ (full line) and the Coulomb form factor
%$f (k)$ (dashed line).}
%\end{figure}
\begin{figure}
\includegraphics[scale=0.4,bb=43 34 600 405,clip=]{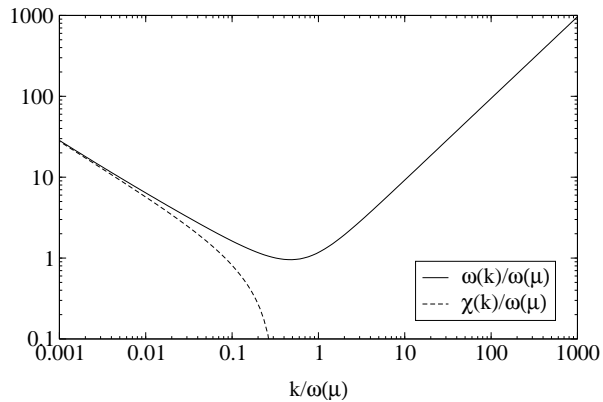}% Here is how to import EPS art
\caption{\label{figure5F5}The gluon energy $\omega (k)$ (full line) and the curvature $\chi (k)$
(dashed line).}%(d\(k\)  and  f\(k\)) show
\end{figure}
\begin{figure} [h]
\includegraphics[scale=0.4,bb=32 36 600 434,clip=]{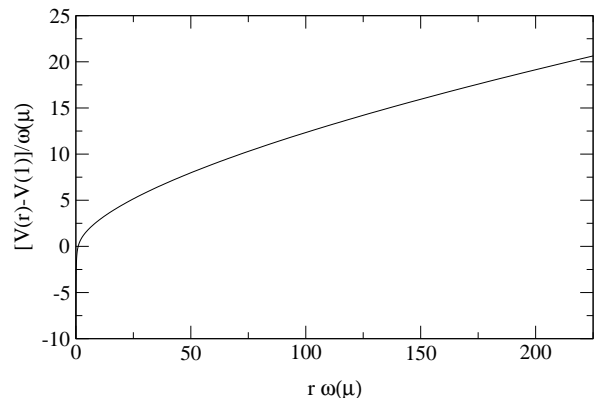}% Here is how to import EPS art
\caption{\label{fig5F6} The static quark potential.}
\end{figure}
%\begin{figure}
%\centerline{
%\label{5F5}
%\epsfysize=7.0cm
%\epsffile{figure5F5.eps}
%}
%\caption{The gluon energy $\omega (k)$ (full line) and the curvature $\chi (k)$
%(dashed line).}
%\end{figure}
%\begin{figure}
%\centerline{
%\label{5F6}
%\epsfysize=6.0cm
%\epsffile{figure5F6.eps}
%}
%\caption{The static quark potential (\ref{F9})}
%\end{figure}
%shows the static quark potential
All these numerical results are in full accord with the ultraviolet and infrared
behaviour extracted above analytically within the angular approximation. 
Finally, fig.5 shows the static quark potential
which is obtained in the present approach as
the expectation value of the last term in the Yang-Mills Hamiltionian, 
eq.(\ref{1}), when the color density $\rho^a$ is identified with 
that of a static quark-antiquark pair. The obtained potential interpolates
between a Coulomb potential at small distances and an (almost) linearly rising
confinement potential at large distances. Numerically we find, that its Fourier
transform diverges for $k \to 0$ as $1 / k^{3.7}$ instead of $1 / k^4$, which
is required for a strictly linear rising potential
\footnote{In ref.\cite{R14} the power $1 / k^{3.6}$ was found.}.
Approximating the confining
potential by a linear one we can fix the energy scale by the string tension. In
ref.\cite{R7} it was found that the string tension in the Coulomb potential,
$\sigma_{coul}$, is by about a factor of three larger than the asymptotic string
tension $\sigma$ extracted from large Wilson loops. Using $\sigma_{coul} = 3
\sigma$ and the canonical value $\sigma = (440 MeV)^2$ one finds that the
minimum in the gluon energy (see fig.4) occurs at $k_{min} \approx 1.4 GeV$, which
is of the order of the glue ball mass. This corresponds to a minimal single
gluon energy $\omega (k_{min})  \approx 3 GeV$. \newline
%\bi
%\no
To summarize, by approximately solving the Yang-Mills Schr\"odinger equation in
Coulomb gauge by means of the variational principle using the trial wave
functional (\ref{2}), which embodies the dominance of the field configurations
on the Gribov horizon, we have been able to describe simultaneously quark and
gluon confinement. A more detailed presentation will be given elsewhere.
\vspace{1.4cm}
\bi

\no
{\it Acknowledgements:}
%\no
%Acknowledgements:
%\bi
%
%\no
The authors are gratefully to R. Alkofer, C. Fischer, O. Schr\"oder, 
A. Szczepaniak and E. Swanson
for useful discussions. This work was supported in part by
Deutsche Forschungsgemeinschaft under contract DFG-Re 856/4-3 and by the
European Graduate College T\"ubingen-Basel.
\bi

\no

\end{document}